\documentclass[sigconf,10pt]{acmart}
\usepackage{tikz}
\usepackage{amsmath}
\usepackage{amsmath}
\usepackage{comment}
\usepackage{xspace}
\usepackage{graphicx}
\usepackage[colorinlistoftodos]{todonotes}
\usepackage{enumitem}
\setlist{nolistsep,leftmargin=2.0mm}
\usepackage{caption}
\usepackage{subcaption}
\usepackage{breakurl}
\usepackage{listings}
\usepackage{filecontents}
\usepackage{titlesec}
\titlespacing*{\section}{0pt}{0.5\baselineskip}{0.2\baselineskip}
\titlespacing*{\subsection}{0pt}{0.1\baselineskip}{0.1\baselineskip}
\usepackage[font={bf,it,footnotesize},labelsep=colon]{caption}
\PassOptionsToPackage{hyphens}{url}
\usepackage{flushend}
\newcommand{\name}{{\sc \bf DeepPlace}\xspace}

\setcopyright{acmlicensed}

\begin{document}
\emergencystretch 3em
\copyrightyear{2019} 
\acmYear{2019} 
\acmConference[APSys '19]{10th ACM SIGOPS Asia-Pacific Workshop on Systems}{August 19--20, 2019}{Hangzhou, China}
\acmBooktitle{10th ACM SIGOPS Asia-Pacific Workshop on Systems (APSys '19), August 19--20, 2019, Hangzhou, China}
\acmPrice{15.00}
\acmDOI{10.1145/3343737.3343741}
\acmISBN{978-1-4503-6893-3/19/08}

\title{DeepPlace: Learning to Place Applications in Multi-Tenant Clusters}


\author[S. Mitra]{Subrata Mitra}
\affiliation{Adobe Research}
\author[S.S. Mondal]{Shanka Subhra Mondal*}
\affiliation{IIT Kharagpur}
\author[N. Sheoran]{Nikhil Sheoran}
\affiliation{Adobe Research}
\author[N. Dhake]{Neeraj Dhake*}
\affiliation{IIT Bombay}
\author[R. Nehra]{Ravinder Nehra*}
\affiliation{IIT Roorkee}
\author[R. Simha]{Ramanuja Simha*}
\affiliation{Oakridge National Lab}
\thanks{\bf{$*$ Work done while at Adobe Research}}
\begin{abstract}
Large multi-tenant production clusters often have to handle a variety of jobs and applications with a variety of complex resource usage characteristics. It is non-trivial and non-optimal to manually create placement rules for scheduling that would decide which applications should co-locate. In this paper, we present \name, a scheduler that learns to exploits various temporal resource usage patterns of applications using Deep Reinforcement Learning (Deep RL) to reduce resource competition across jobs running in the same machine while at the same time optimizing for overall cluster utilization. 
\end{abstract}
\maketitle
\section{Introduction}
\label{sec:introduction}
Today, large production environments often need to handle a large variety of applications, including but not limited to interactive (user-facing) services, latency sensitive applications, batch analytics jobs, stream processing, iterative computations, maintenance services, etc. The standard practice today is to deploy these applications as containers which are then managed by various container orchestration engines such as Docker-Swarm~\cite{docker-swarm}, YARN~\cite{yarn}, Mesos~\cite{mesos}, or Kubernetes~\cite{kubernetes}.
%
%
These orchestration engines allocate resources (e.g., CPU and memory) to these jobs according to the estimated \textit{resource limits} provided by the developers~\cite{kubernetes}. In a multi-tenant shared cluster, if multiple applications compete for the same shared resources, they slow each other down due to resource contention~\cite{maji2015, maji2014, pythia}.
Thus to reduce the chances of contention, orchestration engines use developer specified \textit{affinity}, and \textit{anti-affinity})~\cite{kubernetes, medea_eurosys2018} rules to place applications on different machines.
%
%
%
For stateless-services, resource estimates can be a bit aggressive, such that the resources allocated to each of the deployed containers would be enough to make it run smoothly, while the load fluctuations can be handled through an autoscaling mechanism by increasing or decreasing the number of deployed containers on the fly.
For stateful-services, autoscaling can be really tricky and reactive migration of containers across machines
have high overheads~\cite{medea_eurosys2018}. Hence, the containers are usually deployed with very conservative estimates by specifying large \textit{resource limits} so that they can sustain phases with substantial increase in the resource demands. However, periods with such high resource usages are rare and often span only a very short fraction of the life-cycle of application, leading to resource wastage during the comparatively idle times. 

In most of the real production systems, not all the applications would require to use the peak resource at the same time, and not all phases of their execution would contend for resources in a similar manner~\cite{pythia, atc2018, googlesocc}.
\begin{figure*}[t]
\begin{minipage}[t]{0.28\textwidth}
\includegraphics[width=1.0\textwidth]{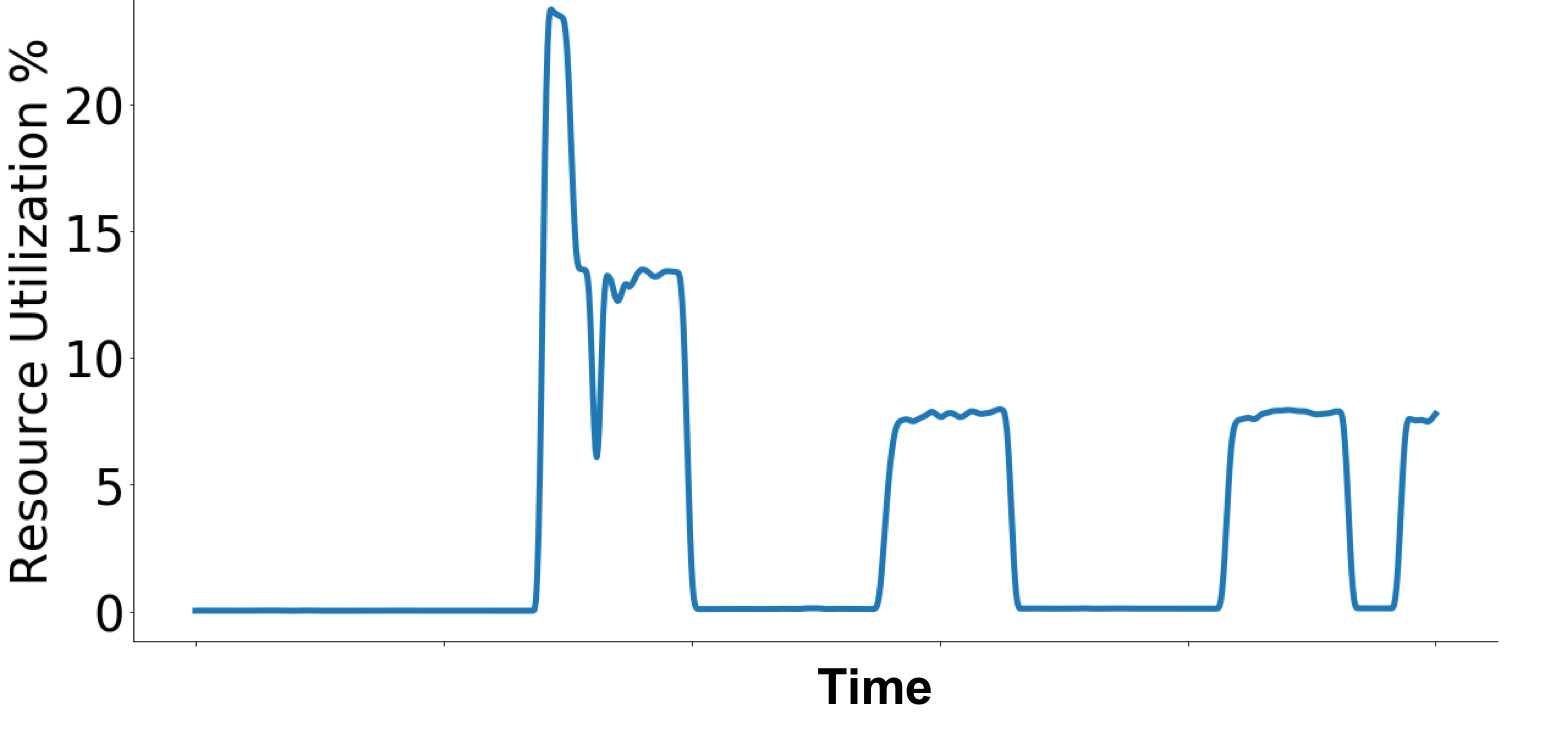}
\end{minipage}
\hfill
\begin{minipage}[t]{0.28\textwidth}
\includegraphics[width=1.0\textwidth]{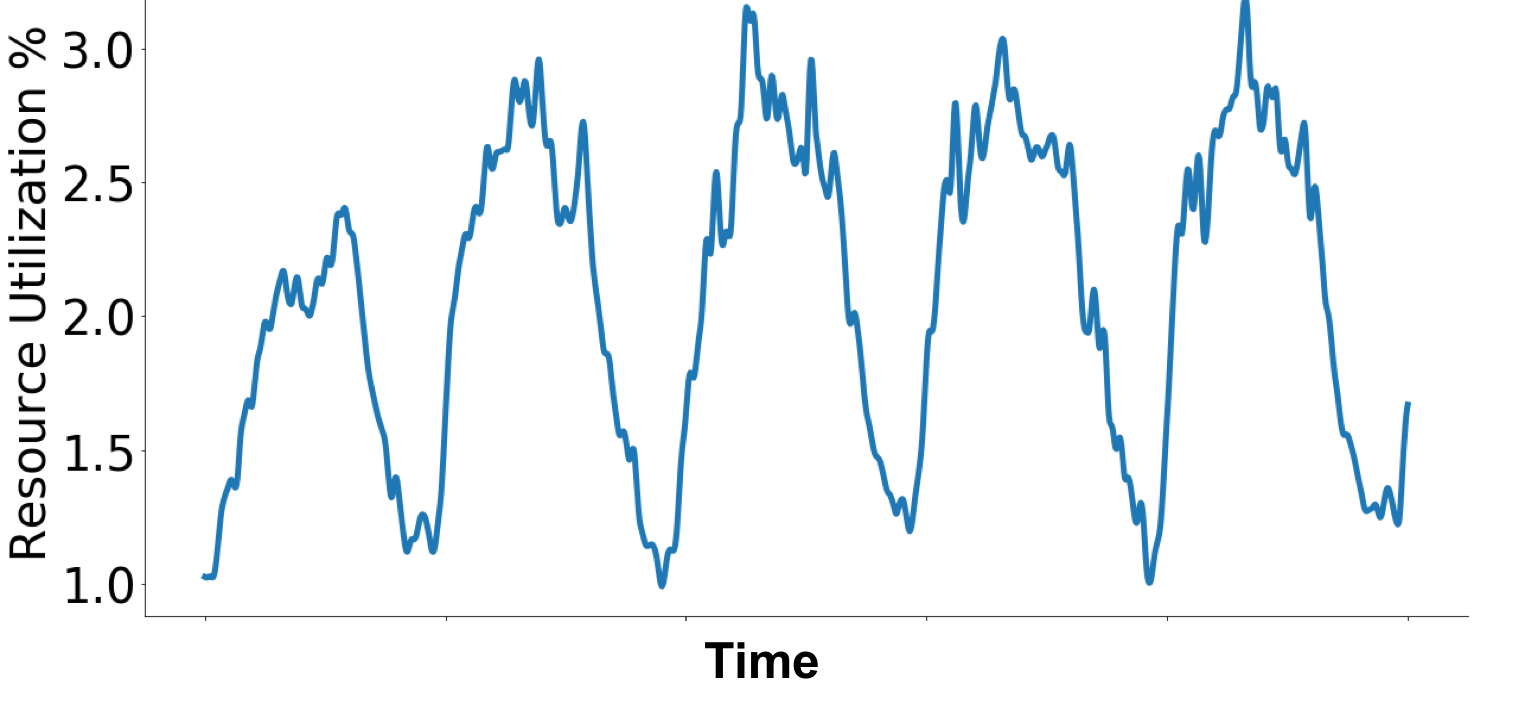}
\end{minipage}
\hfill
\begin{minipage}[t]{0.28\textwidth}
\includegraphics[width=1.0\textwidth]{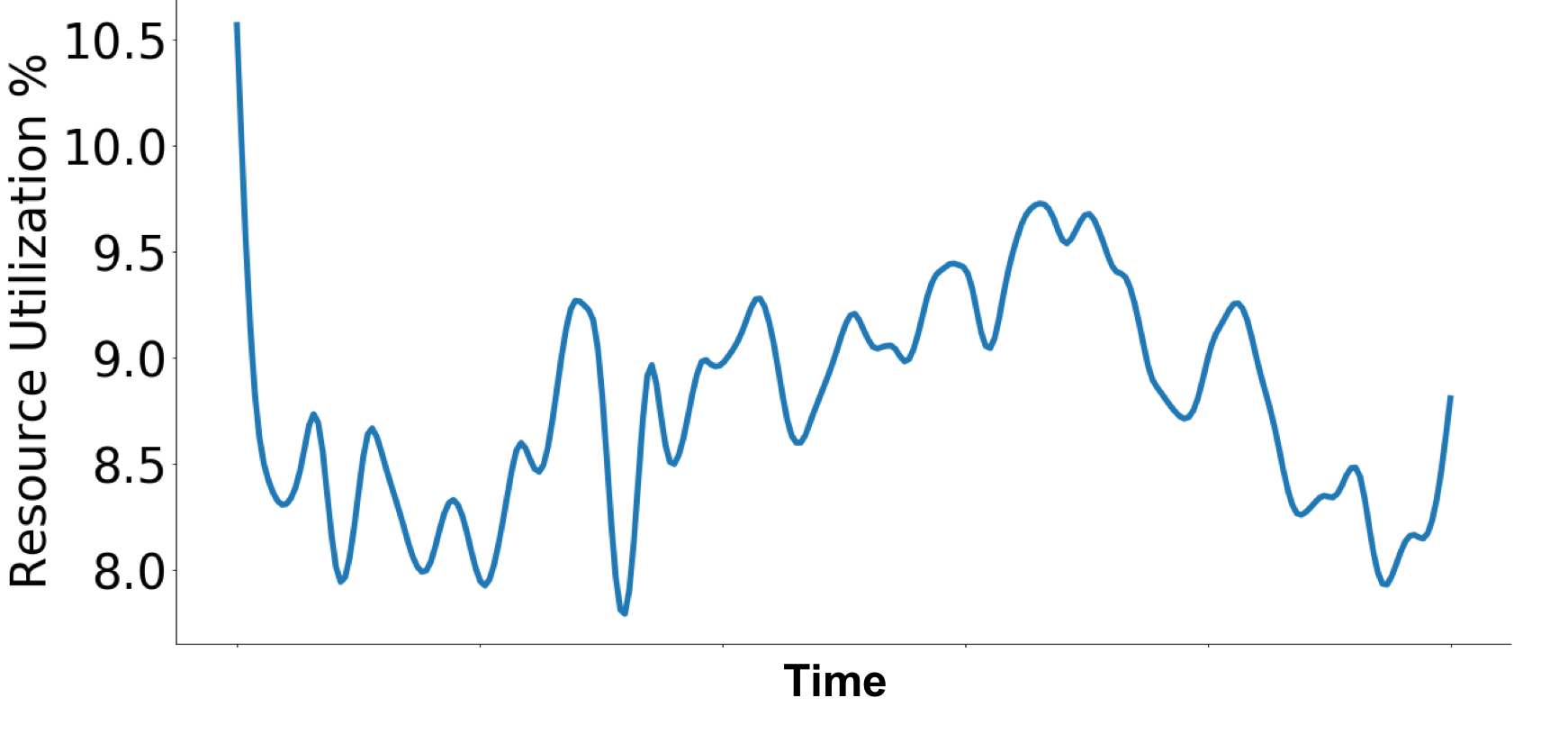}
\end{minipage}
\caption{Examples of resource usages by three production services with several \textit{"peaks"} and \textit{"valleys"}.}
\label{fig:example_usage}
\end{figure*}
For example, Figure~\ref{fig:example_usage} shows the CPU usage characteristics of three production services in our cluster, across various phases of their execution. Their temporal resource usages show several \textit{"peaks"} and \textit{"valleys"}, some more regular than the others.
For user-facing services, such temporal resource usages can have daily and seasonal patterns due to fluctuations in user-demands~\cite{googleworkload} (e.g. some services are mostly used during working hours, while some services are mostly used during major holidays seasons). 
Applications can have different resource usage patterns across algorithmic phases~\cite{opprox, zhang2007, pythia}, e.g., between the \texttt{map} and \texttt{reduce} phases in map-reduce jobs. Because of these temporal variations in the resource usage, \textit{deploying for peak} using developer-provided limits is inefficient from an overall resource utilization perspective. The variety of applications and the complexity of their temporal resource usage patterns makes it infeasible for the developers to express the placement logic in terms of existing placement rules available in
current schedulers, e.g., affinity and anti-affinity rules in Kubernetes~\cite{kubantiaffinity}.

%
%
Borg~\cite{borg} partially addresses this problem by packing a mix of high and low priority jobs in each machine, so that high priority jobs can expand during load spikes whereas low priority jobs can take advantage during the idle periods of the high priority jobs.
However, not all clusters see such a health mix of low priority jobs to effectively fill the valleys of the high priority jobs.

Along with the temporal usage patterns, some jobs might have dependent succeeding jobs that rely on the completion of the first job. These dependencies can be intra or inter services. For example, a customer might have a nightly recommendation model builder, post completion of which a service kicks in to generate a new set of recommendations. A job scheduler that is aware of such dependencies can further utilize this information to efficiently schedule the existing jobs while making room for the upcoming jobs. A central scheduler can even discover \textit{serendipitous dependencies} between different jobs coming from completely different developer groups, opening up scopes for resource alignment among these jobs leading to improved utilization of the cluster.

In this paper, we introduce an early prototype of \name, a self-learning scheduler that can opportunistically place containerized applications such that their temporal resource usages are aligned, resource contentions are minimized, quality of service is maintained and overall utilization improved. 
\name uses deep reinforcement learning (Deep RL) to learn hidden patterns from historical data over time to improve its scheduling policy. 
Essentially, \name treats resource usages of the applications as a \textit{multivariate timeseries} and learns how these timeseries can be placed across different machines so that their resource usages are better aligned. We show through some example cases, how \name can take non-trivial decisions by anticipating future placement requests in order to optimize the overall resource usage in the cluster.
%
For stateful-services, \name helps by minimizing the chances of resource contention, without being overly conservative, leading to operational excellence. For stateless-services, the need for \textit{scaling-up} can be reduced by having a better placement to begin with. With a better placement, a small number of containers might be able to gracefully handle the load up to a certain extent without a need for scaling-up. However, when scaling-up does happen, \textit{where to place} those new additional containers is another crucial question, as usually in container scaling new machines are not spawn off frequently, that can be answered by \name.
\section{Background and Related Work}
\label{sec:background}
\noindent \textbf{Reinforcement Learning.}
%
In RL, at a high-level, an agent interacts with a system and tries to learn an optimized policy.
%
At each timestep $t$, the agent observes the \textit{state} of the system $s_t$, and chooses to take an \textit{action} $a_t$ that changes the state to $s_{t+1}$ at timestep $t+1$, and the agent receives a reward $r_t$. The agent tries to maximize the received \textit{reward} which would help it to learn an optimized policy.
It is assumed that the state transitions and rewards are stochastic and the state transition probabilities 
and rewards depend only on the state of the environment $s_t$ and the action taken by the agent $a_t$ (i.e., show Markov property~\cite{sutton1998reinforcement}).
%

The objective is to maximize the expected cumulative discounted reward: $E[\sum_{t=0}^{\infty}\gamma^{t}r_t]$ where $\gamma \in (0,1)$ determines how much the future rewards contribute to the total reward. More details of theoretical background of RL can be found in
\cite{sutton1998reinforcement} and \cite{deeprm}.
Inspired by the recent trends in Deep Reinforcement Learning (DeepRL), in this paper, we use deep-neural-networks (DNNs) as a function approximator for the placement policy that \name wants to learn. The RL algorithm can perform gradient-descent on the parameters of this DNN so that it can maximize the expected cumulative discounted reward over the actions the RL-agent takes. The gradients are estimated by observing the trajectories of execution that are obtained by following the policy.

RL has been used in variety of scenarios including learning complex games~\cite{mnih2013playing, gibney2016google, lample2017playing}, robotics~\cite{mnih2015human, kaelbling1996reinforcement, kober2013reinforcement}, and very recently for video streaming~\cite{neural-abr-2017}, routing~\cite{boyan1994packet, valadarsky2017learning, mestres2017knowledge}, device placement~\cite{mirhoseini-icml-2017}. But, the application of RL to self-learning schedulers has not been thoroughly explored.
%
%
%

\noindent\textbf{Scheduling.}
To the best of our knowledge, recently proposed DeepRM \cite{deeprm} is the only other self-learning scheduler that also attempts to learn novel scheduling policy using DeepRL. DeepRM has a very simplified view of the cluster and thus comes with several limitations.
\begin{enumerate}
\item 
DeepRM assumes that jobs will always take a fixed amount of resources. 
It does not capture their temporal variations.
%
Users often overestimate resource requirements and there can be a significant difference in resource usage between a peak-load and off-peak loads (Figure~\ref{fig:example_usage}). Thus, ignoring such temporal variations and using the user-specified resource-limits for placement is wasteful and leads to low utilization.
\item 
DeepRM models the resource capacity of the compute cluster as a single monolithic block. It does not have a machine specific view and during its scheduling decisions, it does not try to optimize for the set of job or services to be run together to avoid resource contention.
\item 
The single monolithic view of the total resource capacity of the cluster ignores the impacts of resource fragmentation (i.e., the total amount of available resource in the cluster is more than the requirement of a job, but no single machine has that much available resources left.)
\end{enumerate}

Tetris~\cite{grandl2015multi} is another heuristic-based cluster scheduler that takes into account multiple resource dimensions as well as the alignment of the machine's remaining usage with the job's requirement for packing jobs to the machines.

A large body of work has focused on scheduling data-driven applications, long-running user-facing services, ML-services, etc. on multi-tenant commodity clusters covering various aspects such as fairness of resource sharing~\cite{parkes2015beyond, ghodsi2013choosy, joe2013multiresource, popa2012faircloud, ghodsi2011dominant, grandl2015multi}, tail-latency optimization~\cite{suresh2015c3, ren2015hopper, ferguson2012jockey, leverich_eurosys2014, mace2015retro, bobtail, haque2015few} and how to protect latency sensitive application while improving cluster utilization~\cite{pythia, paragon, quasar, borg, heracles, bubble_flux, q-clouds, omega, apollo}. These are distinct from our work as none of these scheduler attempt to \textit{self-learn} the best scheduling strategy by discovering hidden resource usage characteristics and dependence among applications, along multiple resource dimensions. However, some of the proposed techniques (e.g. cycle-per-instruction~\cite{cpi2}) can be used with our technique to further fine-tune reward/penalty design. 
\section{Design}
\label{sec:design}
%
%

\begin{figure}[t]
    \centering
    \includegraphics[width=0.45\textwidth]{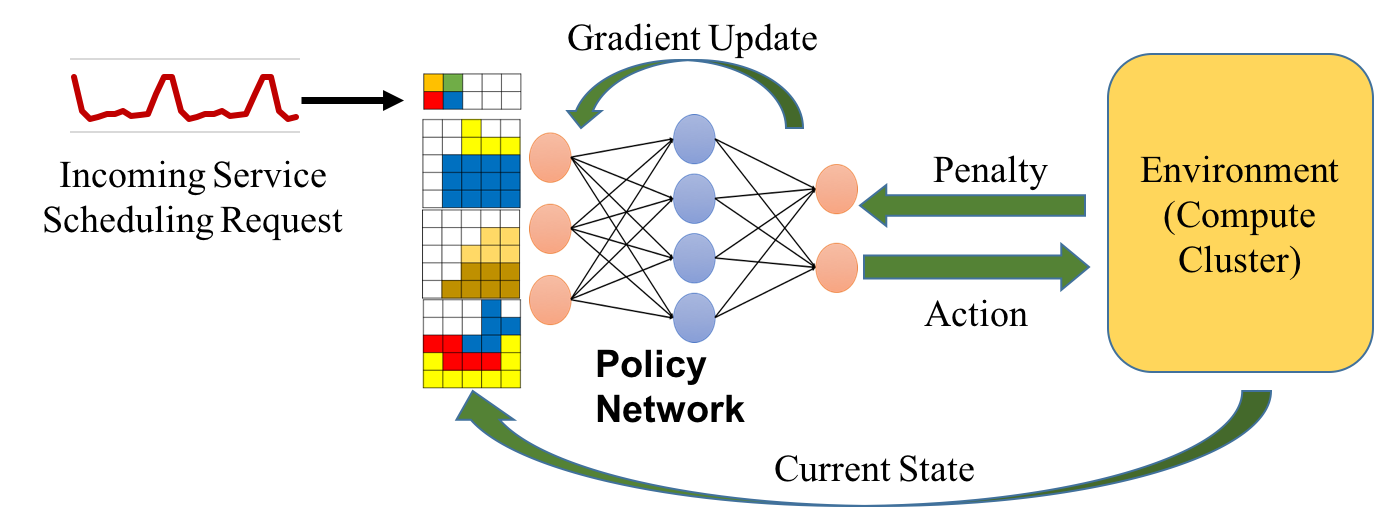}
    \caption{Workflow of \name}
    \label{fig:workflow}
\end{figure}

%
We now describe the design of \name explaining how it operates. \name observes temporal job behavior to optimize its policy, encoded in a DNN-based policy network, using RL.
%
%
%
\name models the scheduling problem as an RL-environment where the compute cluster is composed of $N$ machines on which the application services or jobs are to be scheduled. 
Each such machine has $C_d$ amount of total physical resource capacity for resource dimension $d$ (e.g., CPU, Memory, etc.).
For a job or service $j$, 
\name observes the time-series of the resource usages denoted as 
$r_{d_{j}}$(t), where $r_d$ is the resource usage along the resource dimension $d$. 
\name also keeps track of the current placement map of which services or jobs are running on which machines as well as what are the incoming services or jobs that need to be scheduled in the cluster, as a queue.
The purpose of the queue is to incorporate in the state representation, a view of the upcoming jobs thus allowing the scheduler to learn the arrival patterns and dependencies amongst the jobs. The complete workflow for \name is shown in Figure~\ref{fig:workflow}.

\subsection{State Space Representation}
\begin{figure}[t]
    \centering
    \includegraphics[width=0.45\textwidth]{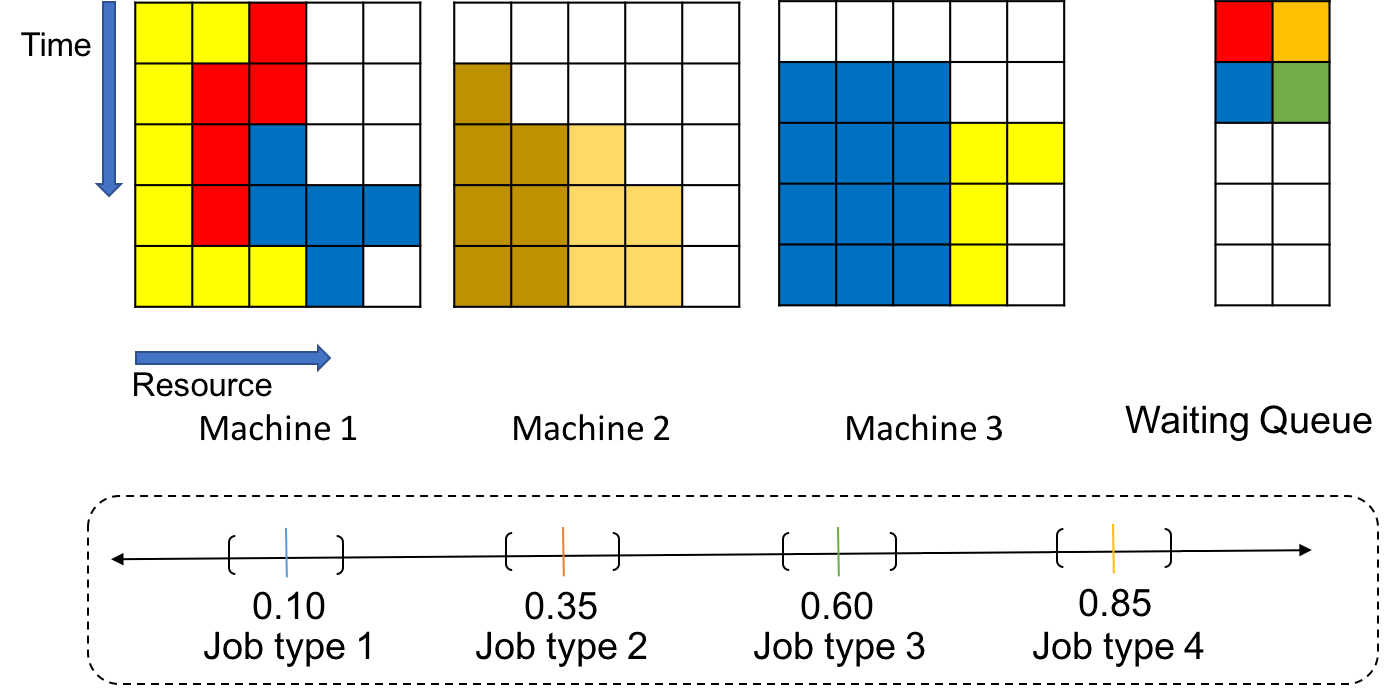}
    \caption{Input space representation of \name}
    \label{fig:state_space}
\end{figure}
\name's state space representation is inspired by~\cite{deeprm}. Though DeepRM's representation for scheduling is designed to answer: \textit{"what job to schedule when"}, \name is designed primarily to answer: \textit{"what job to schedule where"}. 
In extreme cases, \name can delay some scheduling decisions if no suitable placement exists. Thus, \name makes some key improvements in the input-space representation to capture the degree of competition for resources among the jobs sharing the same underlying resources of a machine and their temporal variations in resource usages.

Figure~\ref{fig:state_space} illustrates the input-space representation.
\begin{enumerate}
\item 
State of each machine in the cluster is represented as a 2D matrix or an image with $k$ x $C_{d}$ pixels for each of the resource dimension $d$, where $k$ is the number of previous logical timesteps.
\item Within each machine, the vertical direction of the image (i.e., the matrix) represents the time axis and shows the utilization of jobs for up to $k$ previous logical timesteps, and the horizontal direction represents the amount of resource used by each job/services (quantized into units of resources). 
This type of representation helps the DNN-based RL-agent to learn the temporal resource usage characteristics of each job. $k$ is a configurable parameter that the user can choose. The value of $k$ should be a number reasonably large enough w.r.t. 
scheduling time-scale so that it helps the agent to capture a significant overlap  among applications as well as temporal variations in the resource usage.
However, larger $k$ results in longer convergence time for the RL-agent.

\item For each machine, the number of pixels in the horizontal direction ($C_{d}$) represents the resource capacity of that machine for resource dimension $d$. $C_{d}$ is another configurable parameter that user can choose depending on the granularity of resource usage that needs to be tracked. Larger $C_{d}$ results in longer convergence time for the RL-agent.

\item After each machine, there is a column representing the applications or jobs scheduled and waiting to be run in the machine. This representation is important for \name to take multiple decisions in the same logical timestep. Even if the machine representation is not showing the resource usage of the scheduled application (as time has not proceeded), the column will give an insight to the agent that in the next timestep the application will be running in that machine and hence helps in taking the next action in the same logical timestep.

\item The pixels of the image representing machine-states (i.e., the values in the matrix) are colored differently to denote how much of the
available capacity of the machine is being used at what time by which job.
To make \name scalable, we consider that \name will attempt to learn 
the characteristics of up to $G$ types or equivalence-classes of applications,
and each type of application is represented by a unique number between 0 and 1, both exclusive (this is analogous to a different color of the corresponding pixels in the image).
The unused resources are marked with \textit{white} color (or a value of 0 in the matrix). \name uses these colors (i.e. the numbers) to learn which type of applications when run together can potentially suffer from resource competition and for how long such competition might last.

\item There can be multiple instances of the same application type running in the same machine with an overlap in their duration (e.g. two instances of a face-detection service triggered by two different products). 
These different instances can potentially also create resource contention among themselves (e.g. when the application is highly CPU intensive) and therefore needs to be distinguished and captured by the RL-agent.
We again assign different colors (i.e., floating point numbers) to each instance of the application that are unique but close-by within a small range to the original assigned color for that job type. 

\item \name captures the state of individual machines and combines these machine-level state representations into a cluster-level state representation for creating a holistic input for the policy-network. \name does that by using a trick:
(a) To clearly distinguish between the applications running in different machines, \name adds a different factor for different machines to the number (or color) assigned to the machine usage as well as the column containing the scheduled jobs.
%
For example, if two instances of the same type of application with assigned number representation $0.2$, are running in two different machines (machine 1 and 2), then in the combined state-space representation, these tasks will be represented as \textbf{1.2} and \textbf{2.2} respectively.

\item Along with the combined representation of the machines, \name also keeps a \textit{waiting-queue} in its state-space representation. This queue represents the tasks waiting to be scheduled.
By observing the changes in the queue over time, the RL-agent learns some key dynamics about the arrival characteristics of the jobs, which type of and how many jobs come together, and the temporal dependency amongst them, as previously discussed.
\end{enumerate}

\subsection{Reward/Penalty Design}
\name is driven by negative rewards (penalty) which has the following four components:

\noindent\textbf{Resource contention penalty.} To help \name learn a placement policy that results in better resource alignment (complementary) and avoid resource contention among tasks scheduled in the same machine we use a modified version of cross-correlation to penalize the RL-agent during its learning.
Cross-correlation ($Cr$) is calculated between all pairs of tasks $i$ and $j$ running on the same machine across resource dimension $d$ as follows:

\vspace{-1.0em}
\footnotesize{$$\text{Cr}(i, j, d)= {\sum_{\text{t=0}}^{\text{min(Ti,Tj)} } \text{res\_usage}(i, t, d) \times \text{res\_usage}(j, t, d)} $$}

\vspace{-1.0em}
\normalsize
where $T_i$ is the length of task $i$ and $res\_usage(i, t, d)$ is the instantaneous resource demand across dimension $d$ by task $i$ at time $t$. Cross-correlation formula amplifies the effect of two peaks being scheduled together.
The $Cr$ for a particular state of the cluster is calculated by taking the sum of cross-correlation of each machine, which includes across all the resource dimensions (CPU or memory), the cross-correlation of each task with every other task in that machine.

\noindent\textbf{Resource over-utilization penalty.} To prevent scheduling of more tasks than that can be handled by a machine, there is a high penalty if the machine is not able to meet the resource requirement of tasks scheduled in that machine. It is calculated by adding a high negative factor each time a machine is unable to provide appropriate resources to the running tasks.

\noindent\textbf{Wait-time penalty.} To prevent the scheduler from holding jobs for a long time in search of a better place, we add a constant penalty proportional to the state of the waiting queue. It is equal to the number of waiting tasks in the queue multiplied by a negative constant at each time.

\noindent\textbf{Under-utilization penalty.} Since our goal is to improve overall utilization of the cluster by helping the scheduler learn how to achieve tighter packing and pack on less number of machines, if possible, we add a penalty proportional to the sum of unused resources in the \textit{used machines}. White pixels in our state-representations denote the number of unused resources at any given time.
\section{Implementation}
\label{sec:implementation}
We use the modified version of REINFORCE algorithm as mentioned in~\cite{deeprm}. The policy network 
consists of a single hidden layer of 20 neurons followed by output neurons equal to the number of actions (number of machines under consideration). We use a 36 core CPU server and python multiprocessing to create multiple workers (equal to \textit{batch size+1}) each operating on distinct examples, taking a fixed number of trajectories and accumulating gradients. The last worker is used to combine the gradients of each worker and send to the policy network for updating the parameters. This gives a major improvement in the training speed.
The training time increases significantly as we increase the cluster load. It also depends significantly on the type of applications under consideration (For example, Long running vs Short running jobs).
For the hidden layer, we use Relu activation function, while for the output layer we use softmax activation. We use Adam optimizer and a learning rate of 0.001. The number of trajectories taken by each worker is fixed at 20.
\begin{figure*}[t]
\begin{minipage}[t]{0.3\textwidth}
\includegraphics[width=1.0\textwidth]{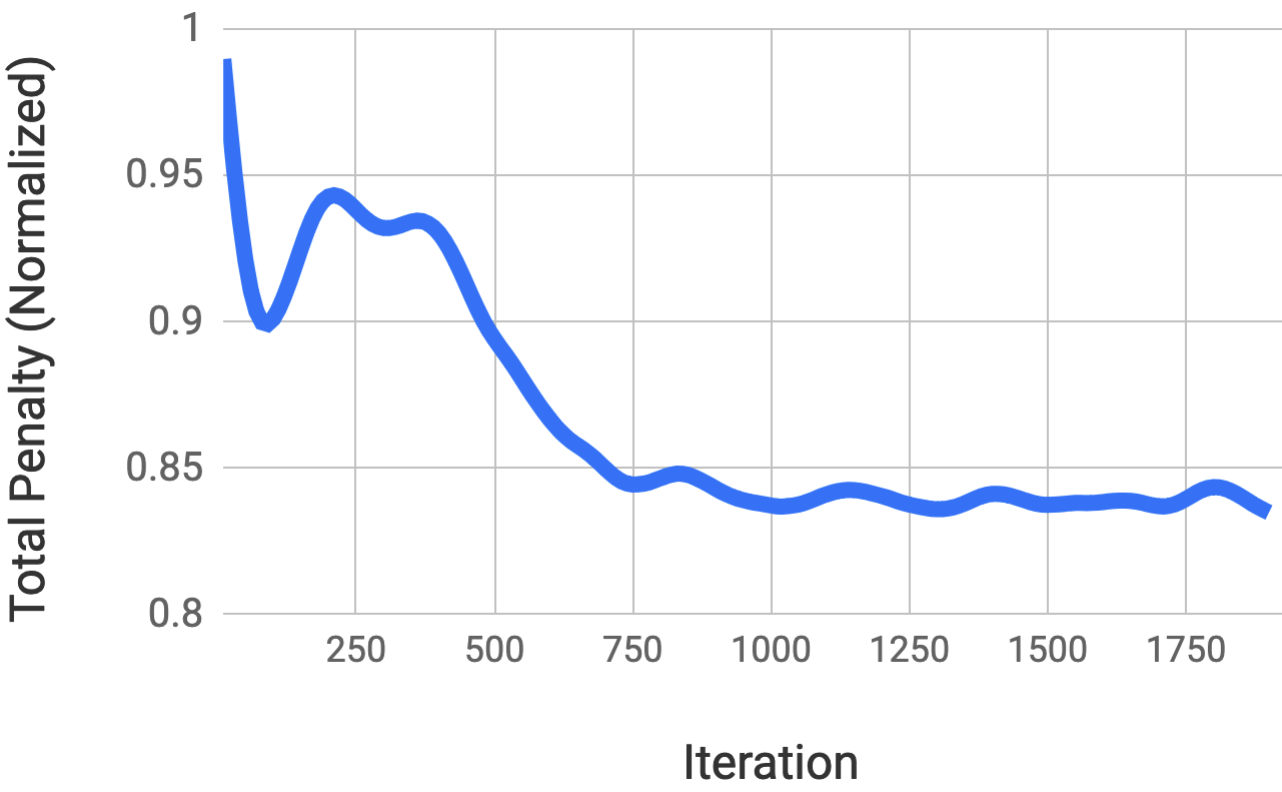}
\caption{Convergence of \name's training under 50\% average load}
\label{fig:convergence}
\end{minipage}%
\hfill
\begin{minipage}[t]{0.3\textwidth}
\includegraphics[width=1.0\textwidth]{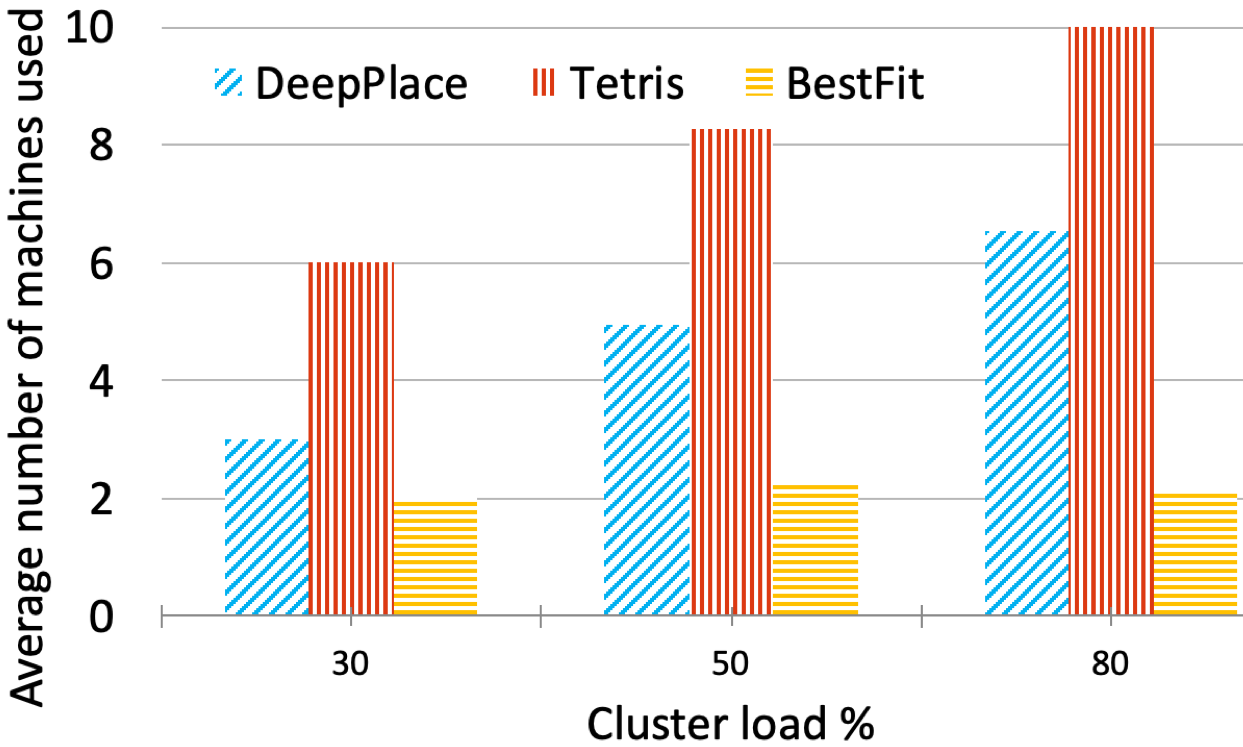}
\caption{Comparison of number of machines used}
\label{fig:machine_used}
\end{minipage}%
\hfill
\begin{minipage}[t]{0.3\textwidth}
\includegraphics[width=1.0\textwidth]{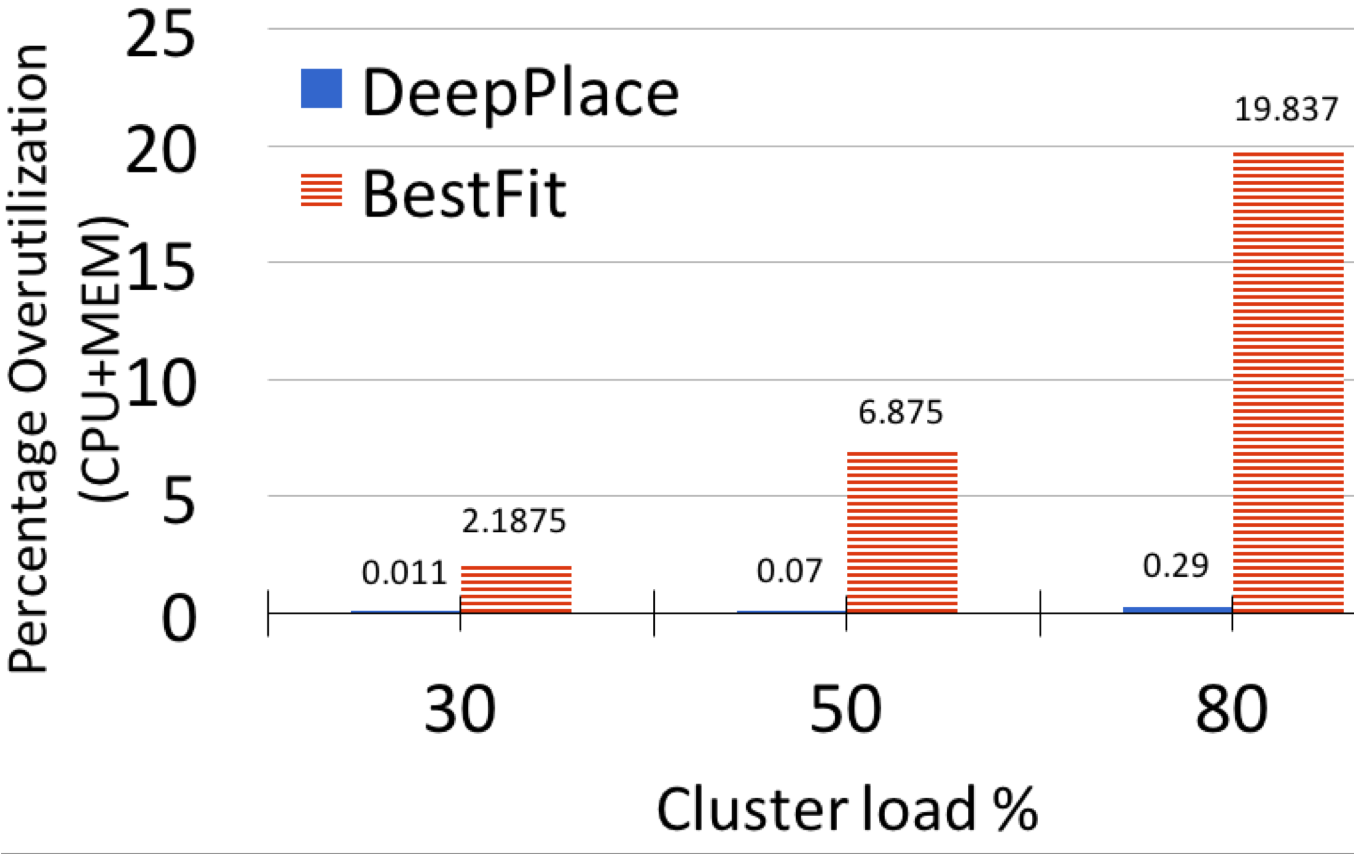}
\caption{Comparison of over-utilization in the cluster}
\label{fig:overshoot}
\end{minipage}
\end{figure*}

\begin{figure*}[t]
\begin{minipage}[t]{0.49\textwidth}
\begin{subfigure}[t]{0.49\textwidth}
\includegraphics[width=1.0\textwidth]{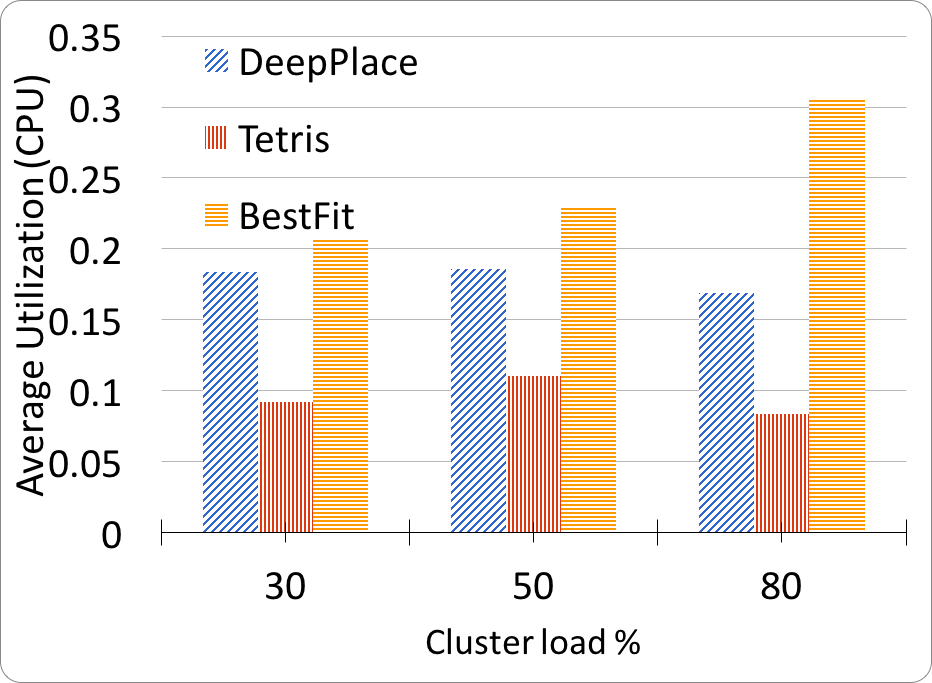}
\caption{CPU utilization}
\label{fig:cpu_util}
\end{subfigure}%
\hfill
\begin{subfigure}[t]{0.49\textwidth}
\includegraphics[width=1.0\textwidth]{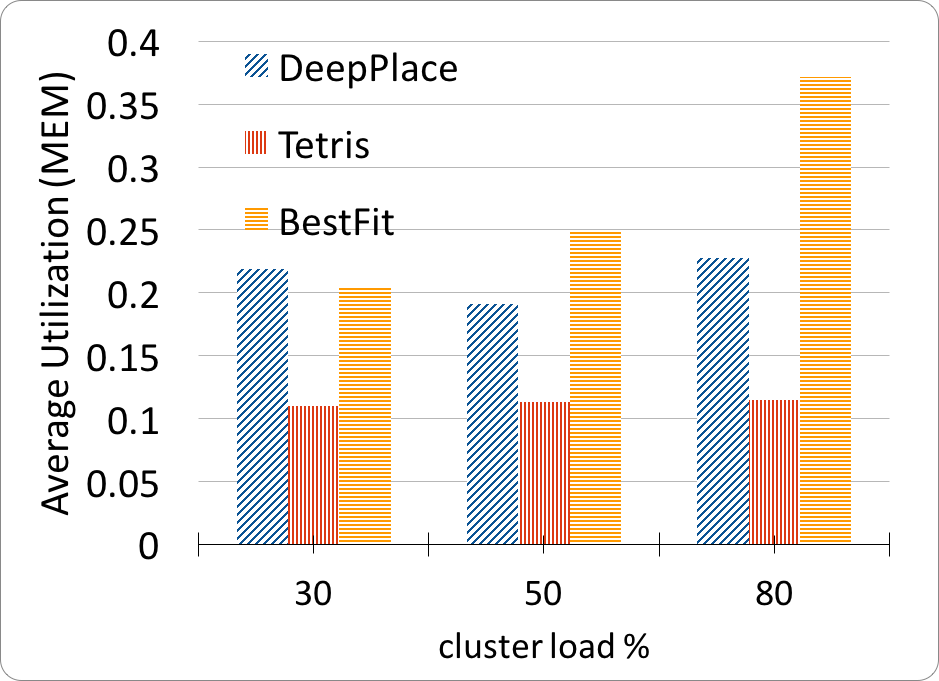}
\caption{Memory utilization}
\label{fig:mem_util}
\end{subfigure}
\caption{Comparison of average resource utilization in the cluster}
\label{fig:util}
\end{minipage}
\hfill
\begin{minipage}[t]{0.49\textwidth}
\begin{subfigure}[t]{0.49\textwidth}
\includegraphics[width=1.0\textwidth]{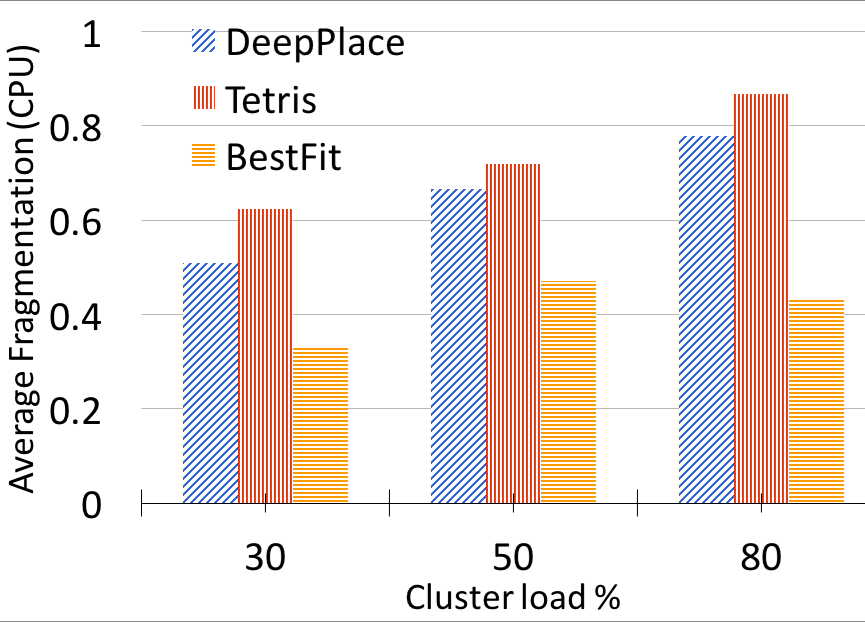}
\caption{CPU fragmentation}
\label{fig:cpu_frag}
\end{subfigure}%
\hfill
\begin{subfigure}[t]{0.49\textwidth}
\includegraphics[width=1.0\textwidth]{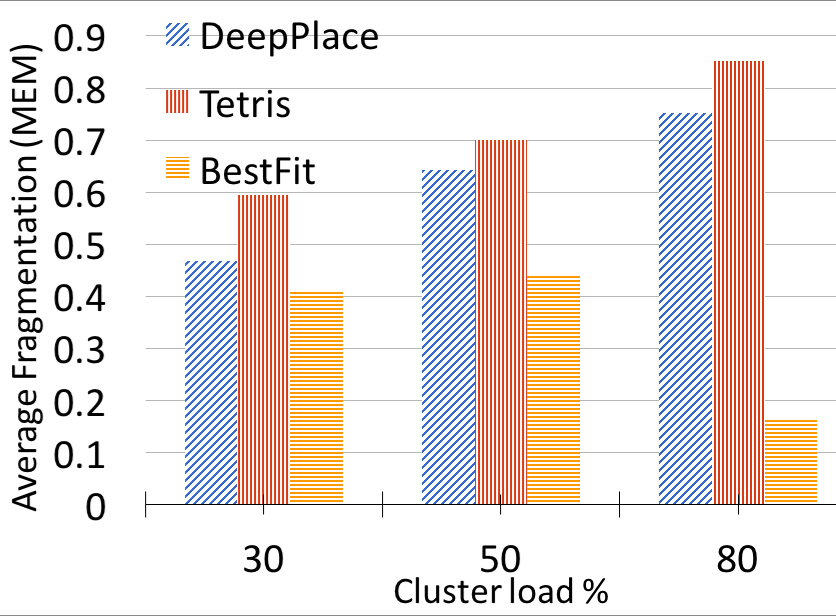}
\caption{Memory fragmentation}
\label{fig:mem_frag}
\end{subfigure}
\caption{Comparison of resource fragmentation level in the cluster}
\label{fig:frag}
\end{minipage}
\end{figure*}

\section{Evaluations}
\label{sec:preliminary}
%
%
\noindent\textbf{[Workload.]} In our evaluation setup, jobs arrive online as a Poisson process. The average job arrival rate is calibrated to create three average cluster load scenarios: 30\%, 50\% and 80\%. 
In our setup, 50\% of the jobs are \textit{long running} and the other half are \textit{short running}. Each job has 2 dimensions of resource requirements: CPU and memory. The capacity of these two resources in each machine is denoted by $\{1r,1r\}$
For each job, dominant resource usage is randomly chosen to be either CPU or memory. The resource usage of the dominant resource is independently chosen from a uniform distribution between 0.3r and 0.5r. The non-dominant resource usage is also independently and uniformly varied between 0.08r and 0.16r. Thus there is no correlation between the CPU and memory usages.
Temporal resource usage for each job varies as a square wave with period uniformly chosen between 0.2t and 0.5t and width as one-fourth of the period, where $t$ denotes the job length.
Total 50 such different jobs are used for training and 18 for testing. Our evaluation runs with a cluster of 10 machines.
%

\noindent\textbf{[Baselines.]} We compare \name with Tetris~\cite{grandl2015multi}, which schedules jobs on machines based on how well job's resource requirement aligns with the machine's available resources balancing preferences for short jobs and packing in a combined score. We also compare it against \textit{Best Fit} heuristic which allocates the job to the machine having the least units of the dominant resource of the job left.

\textit{Note: It is not possible to directly compare \name with \textbf{DeepRM}~\cite{deeprm} because DeepRM only specifies which job to be scheduled next and does not say on which machine it should be scheduled. Thus DeepRM does not have any concepts of competition for resource usage among applications running in the same machine, resource fragmentation and machine-level over-utilization. Thus, a fair comparison with DeepRM with respect to our desired metrics is not possible.}
%
%

\vspace{0.1em}
\noindent\textbf{[Learning Progress.]}
We first show how \name's learning converge across multiple iterations in Figure~\ref{fig:convergence}. It can be observed that roughly after 1000 iterations, \name's policy learning starts to converge and does not see any further significant drop in the normalized penalty.

\vspace{0.1em}
\noindent\textbf{[Improvement in Cluster Utilization.]}
We measure average utilization of machines for each resource as:
 {\footnotesize
$$\text{Avg Util} = \frac{{\sum_{t=0}^{T} \text{Utilization across all machines at time t}}}{{\text{T} \times \text{max(used machines)} \times \text{resource capacity}}} $$}
where $T$ is the length of the observation period. Since the number of machines that are actually being actively used varies over time, in the denominator, we used maximum number of machines used at any point in time to normalize.
%
%

\vspace{0.1em}
\noindent\textbf{[Comparing Scheduling Efficiency.]}
Here in Figure \ref{fig:util}, we show how \name optimizes for cluster utilization for both CPU and memory. We can see that \name can provide a 68-100\% increase in average utilization compared to Tetris across different cluster-load conditions. This is primarily achieved by efficient packing that requires significantly less number of machines to be used compared to Tetris as shown in Figure~\ref{fig:machine_used}. Further, it can be observed that the gap between \name and Tetris in terms of the number of machines required to accommodate the jobs increases with the increase of the cluster load. 
Although it looks like \textit{BestFit} provides even higher utilization because it just packs the jobs into the machines without any knowledge of peak or future resource usages of the jobs and as a consequence, \textit{BestFit} suffers from huge over-utilization of the resources as shown in Figure \ref{fig:overshoot}. On the other hand, over-utilization due to \name's placement decisions are almost negligible. Tetris already includes peak resource usage information in its placement decision thus resulting in no resource over-utilization. 

%

\noindent\textbf{[Improvement in Resource Fragmentation.]}
Fragmentation score of a cluster at a high-level measure what part of all the available resource in a cluster are concentrated.
{\footnotesize
$$\text{Avg Frag} = 1 - {\sum_\text{t=0}^{\text{T} } \frac{\text{max(available space across all machines at t)}}{\text{Sum of available space over all machines at t}}} $$}
The lower the fragmentation score, the higher the ability of the cluster to schedule unanticipated large jobs. Hence, low resource fragmentation in the cluster is a desirable operational property. In Figure~\ref{fig:frag}, we see \name provides 6-13\% reduction in resource fragmentation compared to Tetris. \name's intelligent placement which takes both temporal resource usage characteristics and job arrival patterns leaves bigger room in the machines (i.e. less fragmentation score) to accommodate unanticipated large jobs.
\section{Discussions}
\label{sec:discussions}
\begin{figure}[t]
\begin{subfigure}[t]{0.4\columnwidth}
\includegraphics[width=1.0\textwidth]{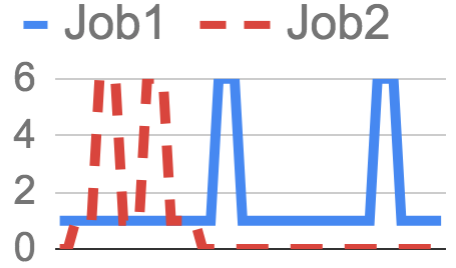}
\caption{Learned example 1}
\label{fig:memory_placement}
\end{subfigure}%
\hfill
\begin{subfigure}[t]{0.4\columnwidth}
\includegraphics[width=1.0\textwidth]{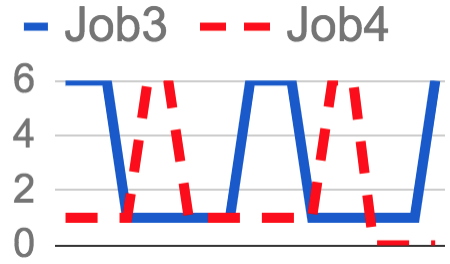}
\caption{Learned example 2}
\label{fig:cpu_placement}
\end{subfigure}
\caption{Examples of learned placement policies}
\label{fig:what_learned}
\end{figure}
In this section, we discuss insights and applicability for real deployments. 

\noindent\textbf{What \name learned?}
Figure~\ref{fig:what_learned} illustrates how \name achieved better packing that ultimately resulted in higher overall utilization.
Figure~\ref{fig:memory_placement} shows how Job1 and Job2 were placed in the same machine because resource intensive parts of Job1 would finish before the resources are required by Job2.
In Figure~\ref{fig:cpu_placement}, resource requirements for Job3 and Job4 alternate in such a manner that they do not exactly overlap with each other and thus were placed in the same machine for better packing. All these patterns were learned by \name on its own without any guiding rules.
%

\noindent\textbf{Scheduling granularity for effectiveness.}
\name looks at \textit{where} to schedule an incoming application so that it can either improve the resource utilization or reduce the resource contention. However, \textit{how often} such a placement decision needs to be made depends on the what kind of workload the cluster is handling. For a cluster handling short or medium-duration batch, cron or interactive applications, frequent placement decisions need to be made and \name can be very useful. On the other hand, for long running services, typically new placement decisions are made less frequently, e.g., when the container for an upgraded service is being deployed, etc. However, if auto-scaling is enabled for these services, taking the decision on where the additional auto-scaled container should be placed in the cluster, can be suggested by \name. 

\noindent\textbf{Cluster size.}
Our input-space representation as well as action-space of the RL is proportional to the number of machines in the cluster. Hence, larger the size of the cluster, the more iterations and training examples it needs for its policy learning to converge. 

\noindent\textbf{Bootstrapping learning in deployments.} 
\name uses historical time-series pattern of resource usages to learn what job is to be scheduled in which machine so that based on their resource usage characteristics, they either improve the overal utilization or avoid aggravating contention by using the same resource at the same time. If \name starts to learn from scratch, it can be long before it sees sufficient examples required for its learning to converge. An option to speed up learning by bootstrapping the RL-agent's policy is by replaying the time-series of historical resource usage through a simulation.


To conclude, in this paper we show an early design prototype of a self-learning scheduler that can exploit the temporal resource usage patterns and arrival dependencies of the jobs to provide a better placement policy and thus achieve better utilization without requiring any manually crafted rules or heuristics.



{\bibliographystyle{acm}
\bibliography{biblio}}

\end{document}